\documentclass[12pt,preprint]{aastex}

\newcommand{\beq}{\begin{equation}}
\newcommand{\eeq}{\end{equation}}

\def\min{{\rm min}}
\def\max{{\rm max}}

\def\ln{{\rm ln}}

\def\3he{$^3$He\,}
\def\he4{$^4$He\,}

\shorttitle{$^3$He and $^4$He Distributions in SEPs}
\shortauthors{Petrosian, Jiang \& Liu}

\begin{document}

\title{Relative Spectra and Distributions of Fluences of $^3$He and $^4$He in
Solar Energetic Particles}

\author{Vah\'{e} Petrosian\altaffilmark{1,2,3}, Yan Wei Jiang\altaffilmark{1,2},
Siming Liu\altaffilmark{4}, George C. Ho\altaffilmark{5} and Glenn, M.
Mason\altaffilmark{5}}

\altaffiltext{1}{Department of Physics, Stanford University, Stanford, CA, 94305
email; vahep@stanford.edu; arjiang@stanford.edu}
\altaffiltext{2}{Kavli Institute for Particle Astrophysics and Cosmology,
Stanford University, Stanford, CA 94305}
\altaffiltext{3}{Also Department of Applied Physics}
\altaffiltext{4}{Los Alamos National Laboratory, Los Alamos, NM, 87545}
\altaffiltext{5}{Applied Physics Laboratory, Johns Hopkins University, 11100
Johns Hopkins Road, Laurel, MD  20723}

\begin{abstract}

Solar Energetic Particles  (SEPs) show a rich variety of spectra and relative
abundances of many ionic species and their isotopes. A long standing puzzle has
been the extreme enrichments of \3he ions. The most extreme enrichments are
observed in low
fluence, the so-called impulsive, events which are believed to be produced at
the
flare site in the solar corona with little scattering and acceleration during
transport to the Earth. In such events  \3he ions show a
characteristic concave curved spectra in a log-log plot.  In two earlier
papers (Liu  et al. 2004 and 2006) we showed  how such extreme enrichments
and such spectra can result in the model developed by  Petrosian \& Liu (2004),
where ions
are accelerated stochastically by plasma waves or turbulence.  In this paper we
address the relative distributions of the fluences of  \3he and  \he4 ions
presented by Ho et al. (2005) which show that  while the distribution of  \he4
fluence (which we believe is a good measure of the flare strength)
 like many other extensive characteristics of solar flare,  is
fairly broad, the \3he fluence is
limited to a narrow range. Moreover, the ratio of the fluences  shows a strong
correlation with the \he4 fluence. One of the predictions of our model presented
in the 2006 paper
was presence of steep variation of the fluence ratio with the level of
turbulence
or the rate of acceleration. We show here that this feature of
the model can reproduce the observed distribution of the fluences with very few
free parameters. The primary reason for the success of the model in both fronts
is because fully ionized \3he ion,  with its unique charge to mass ratio,
can resonantly interact with more plasma modes and accelerate more readily  than
\he4. Essentially in most flares, all
background $^3$He ions are accelerated to few MeV/nucleon range, while this
happens for $^4$He ions only in very strong events. A much smaller fraction of
$^4$He ions reach such energies in weaker events.
\end{abstract}


\section{INTRODUCTION}
\label{intro}
Solar flares are excellent particle accelerators. Some of these particles on
open field lines are observed as solar energetic particles (SEPs) at one AU or
produce type III and other radio radiation. Those on closed field lines can be
observed by the radiation they produce as they interact with solar plasma and
fields. Electrons produce nonthermal bremsstrahlung and synchrotron photons in
the hard X-ray and microwave range, while protons (and other ions) excite
nuclear lines in the 1 to 7 MeV range or may produce higher energy gamma-rays
via $\pi^0$ production and its decay. It appears that stochastic acceleration
(SA) of particles by plasma waves or turbulence  plays an important role in
production of high energy particles and 
consequent plasma heating  in solar flares (e.g., Ramaty 1979; M\"{o}bius et 
al. 1980, 1982;  Hamilton \&  Petrosian 1992; Miller et al. 1997; Petrosian \&
Liu 2004, hereafter PL04).
This theory was applied 
to the acceleration of nonthermal electrons (Miller \& Ramaty 1987;  Hamilton \&
Petrosian 1992). It appears that it can produce many of the observed radiative
signatures such as broad band spectral features (Park, Petrosian \& Schwartz
1997; PL04) and the commonly observed hard X-ray emission from the tops of
flaring loops (Masuda et al. 1994;  Petrosian \& Donaghy 1999). 
It is also commonly believed that the observed relative abundances of ions in
SEPs
favor a SA model (e.g. Mason et al. 1986 and Mazur et al. 1992). More recent
observations have
confirmed this picture (see Mason et al. 2000, 2002, Reames et al. 1994 and
1997, and Miller 2003).
One of the most vexing problem of SEPs has been the  enhancement of \3he
in the so-called {\it impulsive  or \3he-rich events}, which sometimes can be
$3-4$ orders of
magnitude above the photospheric value%
\footnote{In addition there is charge-to-mass ratio dependent enhancement
relative to the photospheric values  of heavy ions in SEPs , and  in few flares
gamma-ray line emissions also points to anomalous abundance pattern of the
accelerated ions (Share \& Murphy 1998; Hua,  Ramaty \& Lingenfelter 1989). We
will not be dealing with these anomalies in this paper.}.
There have been many attempts to  explain this enhancement.
Most of the proposed models, except the  Ramaty and Kozlovsky (1974) model based
on  spalation (which has many problems), rely on resonant wave-particle
interactions and the unique charge-to-mass ratio of \3he  (see e.g. Ibragimov \&
Kocharov 1977; Fisk
1978; Temerin \& Roth 1992; Miller \& Vi\~nas 1993; Zhang 1995; Paesold,
Kallenbach \& Benz 2003). Most  of these model assume presence of some
particular kind of waves which preferentially \underline{heats} \3he ions to a
higher temperature than \he4 ions, which then become
seeds for
subsequent acceleration by some (usually) unspecified mechanism (for more
detailed discussion see Petrosian 2008). None
of these earlier works did a compare  model spectra with
observations.

\begin{figure}[htb]
\begin{center}
\includegraphics[height=4.7cm]{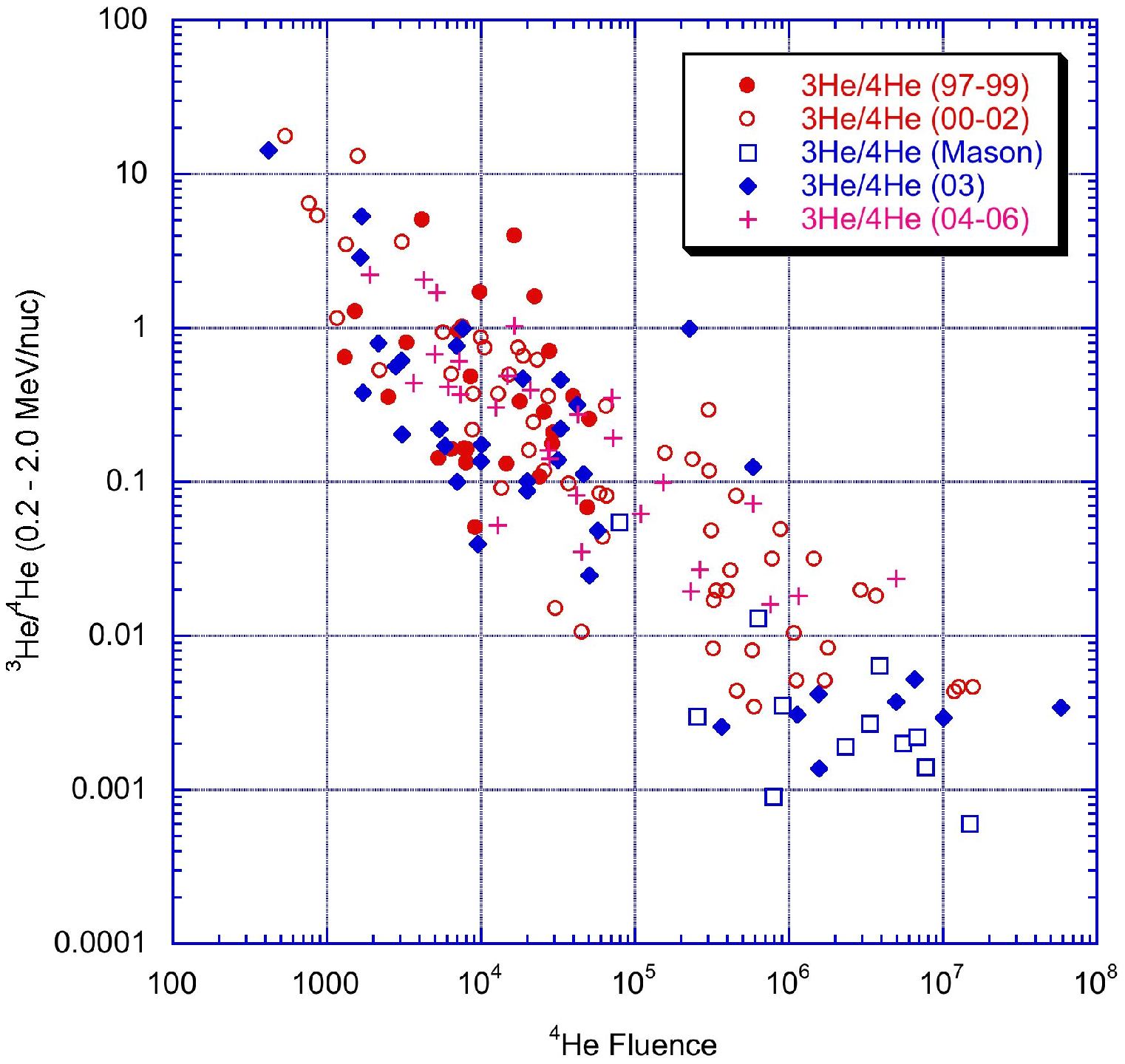}
\includegraphics[height=5.5cm]{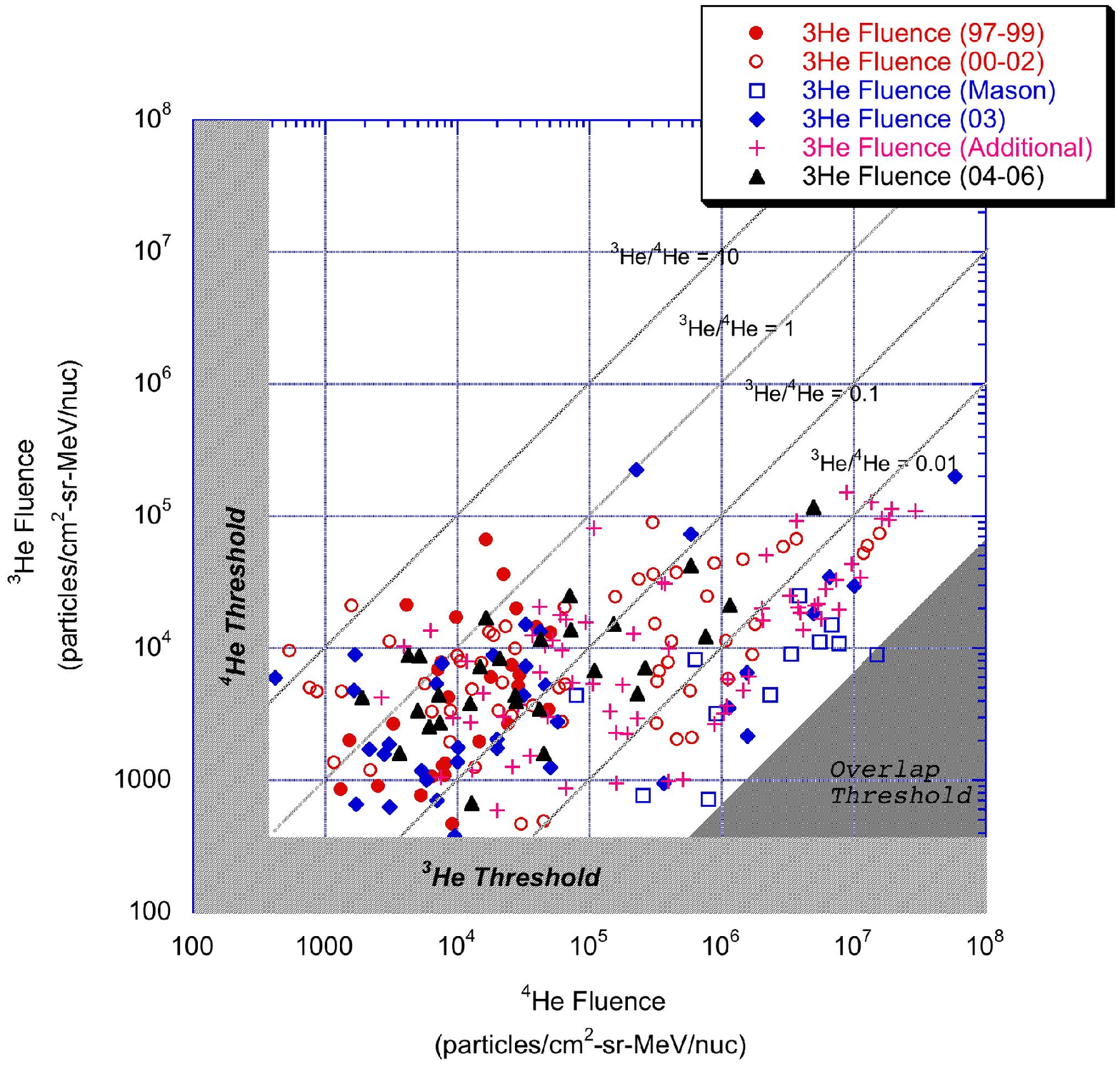}
\hspace{-1cm}
\includegraphics[height=5.9cm]{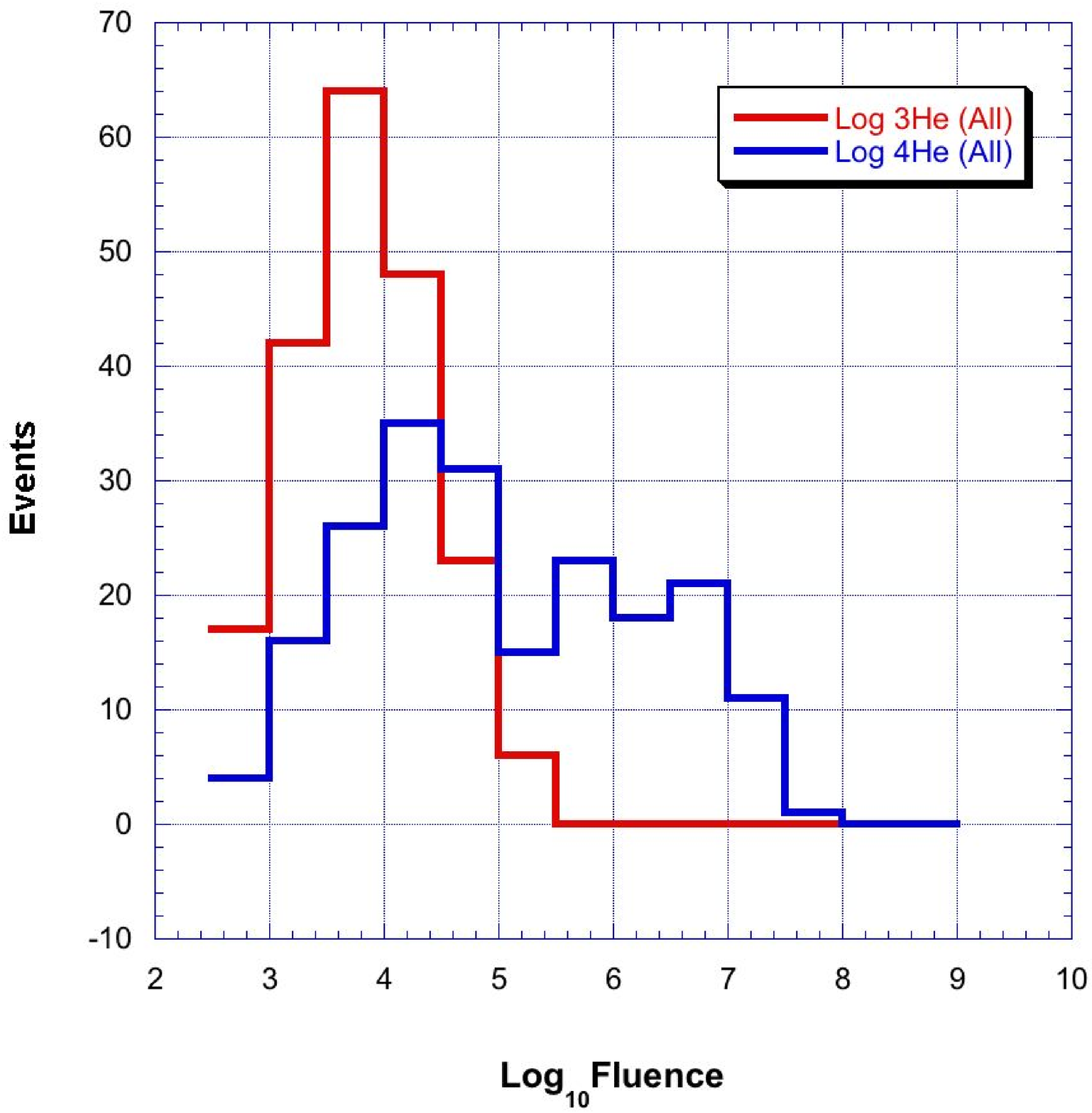}
\end{center}
\caption{\scriptsize
{\it Left:} Variation of the ratio of \3he to \he4 fluences with the fluence of
\he4 showing a continuum of enrichments and a strong anti correlation. {\it
Middle:}  \3he vs \he4 fluences showing a much larger range for the latter while
the former seems to be limited to a small range. Note that the \3he fluences do
not concentrate at the lower end which would be the case if observational
threshold  was affecting their distribution.  {\it Right:} The distribution of
fluences of \3he and \he4. Note that the high end of the \he4 distribution may
be truncated because of the threshold of the fluence ratio (missing point in the
lower left triangle of the middle panel) [From Ho05].  The fluences are in units
of particles/(cm$^{2}$ sr MeV/nucleon).
}
\label{obs}
\end{figure}

In two more recent papers Liu, Petrosian \& Mason 2004  and 2006 (LPM04, LPM06)
have demonstrated that a SA model by parallel propagating waves
can explain both the extreme enhancement of \3he
and can  reproduce the observed \3he and \he4 spectra. In LPM06 it was shown
that the relative fluences of these ions, and to a lesser extent their spectral
indexes, depend on several model parameters so that in a large sample of events
one would expect some dispersion in the distributions of fluences and spectra. 
Ho et al. (2005; Ho05) analyzed a large sample of events and provide
distributions of \3he and \he4 fluences and the correlations between them. Our
aim here is to explore the possibility of explaining these observations
by the above mentioned dependence of the fluences on the model parameters.  In
particular we would like to explain the observations reproduced in Figure
\ref{obs} which shows a strong anti correlation of \3he/\he4 ratio with
\he4 fluence (left panel), but shows essentially no correlation between the two
fluences (middle panel). More strikingly,  the \3he fluence distribution appears
to be relatively narrow and follows a log-normal distribution, while \he4
distribution is much broader and  may have a power law distribution in the
middle of the range, where the
observational selection effects are unimportant. Often
the SEPs are divided into two classes; impulsive-high enrichment and
gradual-normal abundance classes. However, as evident from the left panel of the
above figure there is a continuum of enrichment extending over many orders of
magnitude.%
\footnote{The \he4 distribution shows a weak sign of bi modality but this is not
statistically significant. In this paper we will ignore this feature.}

In the next section we describe some of the model characteristics that can
explain these observations and in \S3 we compare the model predictions with the
observations, specifically the distributions of the fluences. A brief summary
and conclusion is given in \S4.

\section{MODEL CHARACTERISTICS}
\label{model}

The model used in LPM04 and LPM06 which successfully described the enrichment
and
spectra in several flares has several free parameters.  As usual we have the
plasma parameters
density $n$, temperature $T$ 
and magnetic field $B_0$. It turns out that the final results are insensitive to
the temperature  as long as it is higher than $2\times 10^6$K (see Fig.
\ref{ratios} below), which is the case for
flaring coronal loops. It also turns out that only a combination of density  and
 magnetic field ($\sqrt n/B_0$)  comes into play. We  express this as the
ratio of plasma to gyro-frequency of electrons,
$\alpha=\omega_{pe}/\Omega_e$ which is related to the Alfv\'en velocity in unit
of speed of light;
$\beta_A=\delta^{1/2}/\alpha$, where $\delta=m_e/m_p$ is the ratio of the
electron to proton masses. So in reality  we have only one effective free plasma
parameter $\alpha$ or $\beta_A$.  On the other hand, several parameters are
required to describe the spectrum of the turbulence.
Following the above papers we assume broken power laws for the two
relevant modes, the proton cyclotron (PC) and He cyclotron (HeC),
with an inertial range $k_\min<k<k_\max$, and similar power law indexes $q$ and
$q_h$ in
and beyond the inertial range, respectively.%
\footnote{In LPM06 we also have an index $q_l$ describing the power law below
the inertial range which is of minor consequence. For all practical purposes we
can assume a sharp cutoff below  $k_\min$ which means
$q_l\rightarrow \infty$.}
The only difference between the two branches is that the 
wave numbers  $k_\max$ and $k_\min$ for the PC mode are two times higher 
than those for the HeC mode. Finally there is the most important parameter
related
to the total energy density of  turbulence,  ${\cal E}_{\rm tot}$, which
determines both the rate of acceleration and, when integrated over  the
volume of the source region, determines the intensity or the strength of the
event. This parameter is  the characteristic time
scale $\tau_p$ or its inverse the  rate defined as  (see, e.g. Pryadko \&
Petrosian 1997) 
\begin{equation}
\tau_p^{-1} = {\pi\over 2}\Omega_e\left[{4\overline{\cal E}_0\over
B_0^2/8\pi}\right]\,\, \ \ \ \ {\rm with} \ \ \ \ \overline{\cal E}_0 =
{(q-1){\cal E}_{\rm tot}\over 
(k_\min c/\Omega_e)^{1-q}},
\label{taup}
\end{equation}  
for each mode. The factor of 4  arises from having two branches (PC and HeC)
and  two propagation directions of the waves (see LPM06 for details).

As shown in LPM04 and LPM06 papers the main difference between the acceleration
process of \3he and \he4 is in the difference between their acceleration rate or
timescales ($\tau_a$). The other relevant timescales, namely the loss
($\tau_{\rm loss}$) and escape ($T_{\rm esc}$) times are
essentially identical for the two ions (e.g. see left panel of  Fig. 7 of
LPM06). The
acceleration timescales are different mainly at low energies (typically below
one MeV/nucleon), where the acceleration time of \he4 is a longer (by one  to
two orders of magnitude). As
a result at these low energies the \he4 acceleration time may be comparable or
longer than the loss time which makes it difficult to accelerate \he4 ions. Most
of \he4 ions are piled up below some energy (roughly where $\tau_a=\tau_{\rm
loss}$) and
only a few of
them accelerate into the observable range (e.g. see right panel of  Fig. 7 of
LPM06).
However, because the acceleration times
scale as $\tau_p$  while the loss time does not, for higher level
of turbulence (larger $\overline{\cal E}_0$), the acceleration time may fall
below the
loss time so that \he4 ions can be then accelerated more readily (see Fig.
\ref{spectra} below). On the other
hand,
essentially independent of values of any of the above parameters, the \3he
acceleration time at all energies, in particular at low energies, is always far
below its loss time so that in all cases   (except for very high densities or
very low values of $\tau_p^{-1}$)
\3he ions are accelerated easily to high energies. The relative values of the
escape and acceleration times (for
both ions) determine their  high energy spectral cutoffs.  

\begin{figure}[htb]
\begin{center}
\includegraphics[height=5.0cm]{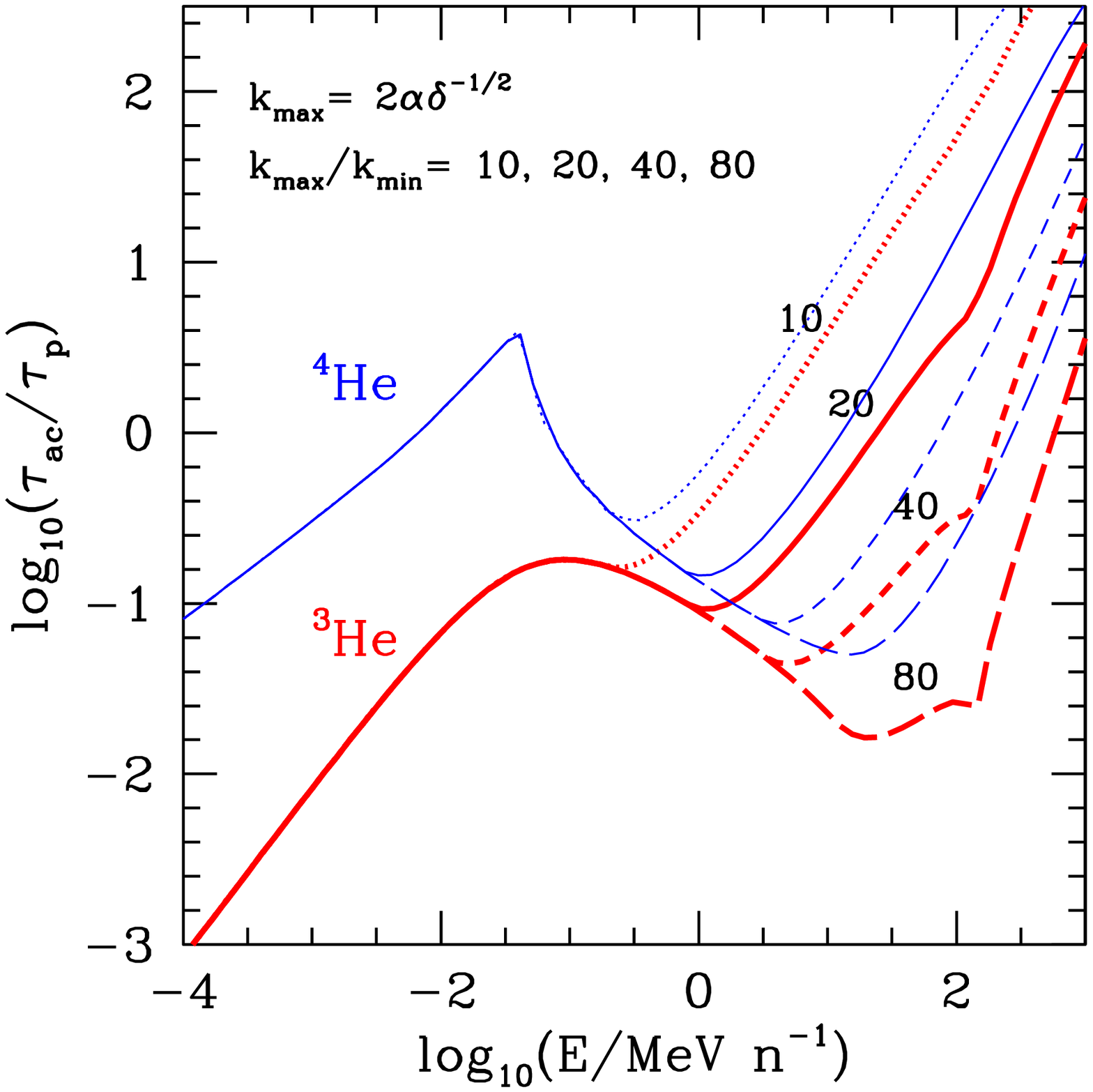}
\hspace{-0.6cm}
\includegraphics[height=5.0cm]{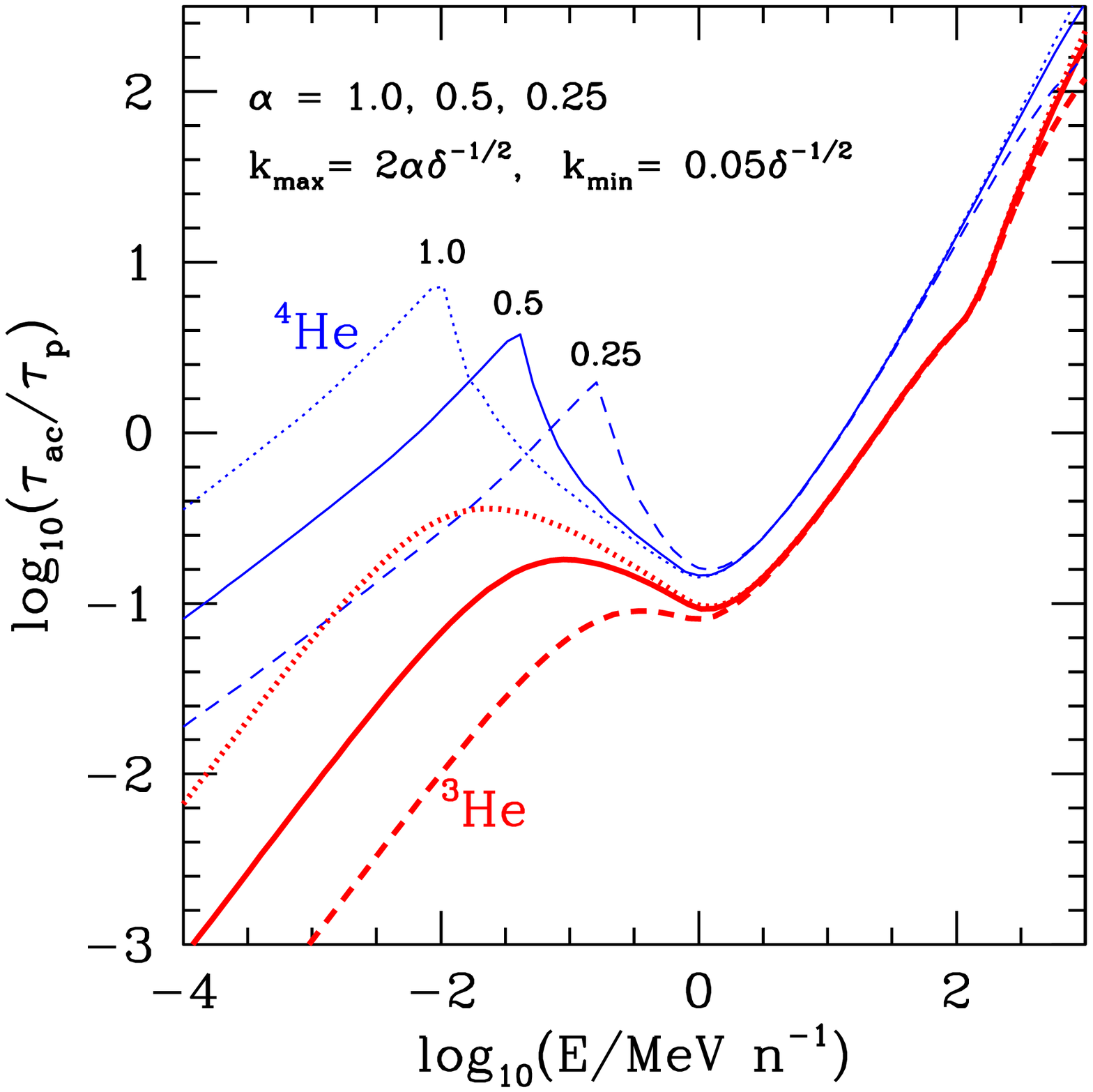}
\hspace{-0.6cm}
\includegraphics[height=5.0cm]{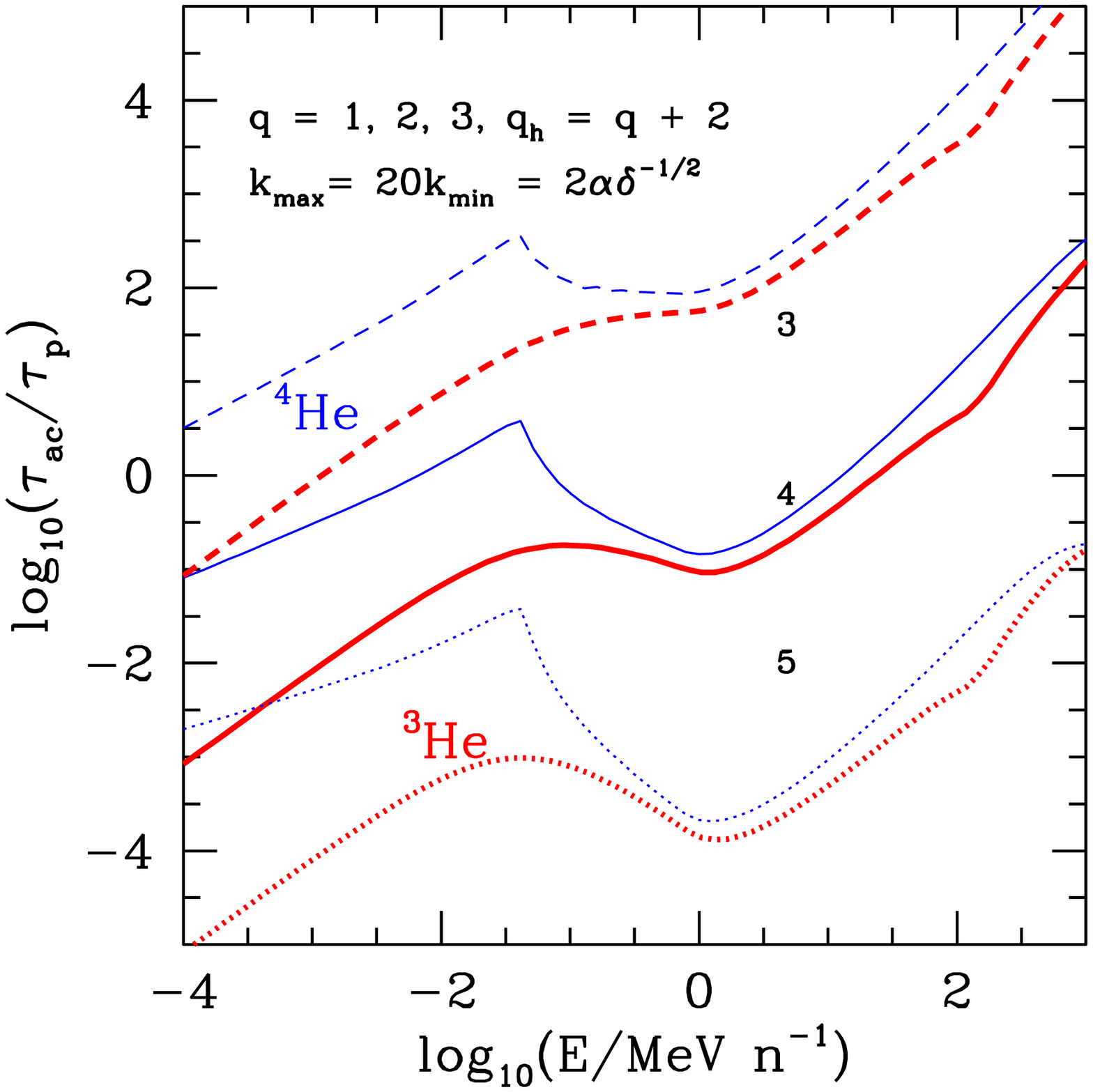}
\end{center}
\caption{\scriptsize
Dependence of the acceleration time of \he4 (thin, blue) and \3he (thick, red)
on $k_\min$ (left),
$\alpha$ (middle) and $q$ (right).  The lines are labeled with the corresponding
numbers of each parameter.
In each case  the solid lines are for the  fiducial model with
$\alpha=0.5,
k_\max=2\alpha\delta^{-1/2}=10k_\min, q=2$ and $q_h=4$. 
} 
\label{acctimes}
\end{figure}

Figure \ref{acctimes} shows variation with energy of acceleration times of \3he
(thick lines) and  \he4 (thin lines) and their dependence on  parameters
$k_\min$, $\alpha$ and  $q$. The remaining parameters $q_h$ and $k_\max$ only
affect the slope of the low energy
end of \he4 which does not affect the spectra noticeably.
It is evident that  the general behavior of the acceleration time scales  described above
(consisting of a low and a high energy monotonically
increasing  branches with a declining transition in between)
is  present in all  models. These features
change only quantitatively and often by  small amounts.  As expected lowering
$k_\min$ decreases the acceleration times at the  high energy branch (left
panel). This is because the lower $k_\min$ waves interact  resonantly with
higher energy ions.  On the other hand, a lower value of $\alpha$ (or larger
Alfv\'en velocity or magnetization) decreases
the times at the low energy branch  (middle panel). Steeper spectra in the
inertial range,
produce a higher rate of acceleration (larger ${\cal E}_0$; see eq.
[\ref{taup}]) and  decrease the overall acceleration time scales (right panel) 

Note  that in this and subsequent figures, $k_\max$ is in units
of $\Omega_p/c$ so that  $k_\max=2\alpha\delta^{-1/2}$ in the labels means an
actual $k_\max=2\Omega_p/v_A= \sqrt{2\beta_p}/r_{g,p}$, where
${\beta_p}=2(v_{th,p}/v_A)^2$ is the plasma beta, and $v_{th,p}=k_BT/m_p$ and
$r_{g,p}=v_{th,p}/\Omega_p$ are the proton thermal velocity and gyro radius. The
scale of $k_\max$  is clearly beyond the MHD regime (where the wave frequency 
$\omega=v_A k\ll \Omega_p$), but is below the proton gyro radius for the chosen
parameters ($\sqrt{2\beta_p}\sim0.03$). 

Using these acceleration rates we calculate spectra of  the two ions (as  in
LPM04 and LPM06) for a range of 
parameters. Figure \ref{spectra} shows three sets of spectra where we
vary $k_\min, \alpha$ and $\tau_p^{-1}$. In each panel the sold lines are for
the fiducial model ($\alpha=0.5, k_\max=2\alpha\delta^{-1/2}=10k_\min, q=2$ and
$q_h=4$) chosen to fit the spectra observed by {\it ACE/ULEIS} for 30 Sep. 1999
event.
The spectral variations here reflect
the above  described variations of the acceleration timescales.  Lower $k_\min$
(or larger inertial range)  yields a larger tail for both ions (left panel).
Variation of $\alpha$ has a similar and smaller effect on \3he spectra but it
affects the \he4 spectra dramatically; for $\alpha\sim 1$ essentially there is
no \he4 acceleration but the  $\alpha\sim 1/4$ model accelerates a large number
of \he4
ions beyond 0.1 MeV/nucleon and into the observable range (middle panel). This
effect is even more pronounced for increasing values of $\tau_p^{-1}$, where a
factor of few increase in the general rate of acceleration (or the level of
turbulence) causes a large increase of the fluence of \he4  (right panel),
because, as stated above, its acceleration time  becomes shorter than its loss
time even al low energies.  All
these spectra show the same general characteristic features. While most \3he
ions are accelerated to  high energies for essentially all  model parameters
appropriate for solar coronal conditions and reasonable level of turbulence,
\he4
ions show a characteristic lower energy bump with a nonthermal hard tail. In
general, the lower energy bump is below the observation range except for low
$\alpha$ and high values of $\tau_p^{-1}$. Since a high level of turbulence is
expected for brighter and stronger events, this means that we get smaller
\3he/\he4 flux or fluence ratios for brighter events. Note that the spectra in
such cases may not agree with observations but this is not troublesome, because
as is well established, the stronger events (the so-called gradual events) are
associated
with CMEs and shocks which most likely will modify the above spectra which are
those of ions escaping the corona. Thus, the higher energy bumps in the spectra
shown here should be considered as seeds for
such further acceleration during the transport from the lower corona to the
Earth, which becomes more likely, and is
expected to change the above spectra more significantly,  for more energetic
events.  Thus if we give up the idea that there are two distinct classes of SEPs
(impulsive and highly enriched and gradual and normal abundance) but that there
is a continuum of events, which observations in Figure \ref{obs} show, then the
above scenario implies that the main acceleration occurs in the solar corona.
Subsequent interactions in CME shocks mainly modify the seed population escaping
the turbulent coronal site.

\begin{figure}[htb]
\begin{center}
\includegraphics[height=5.0cm]{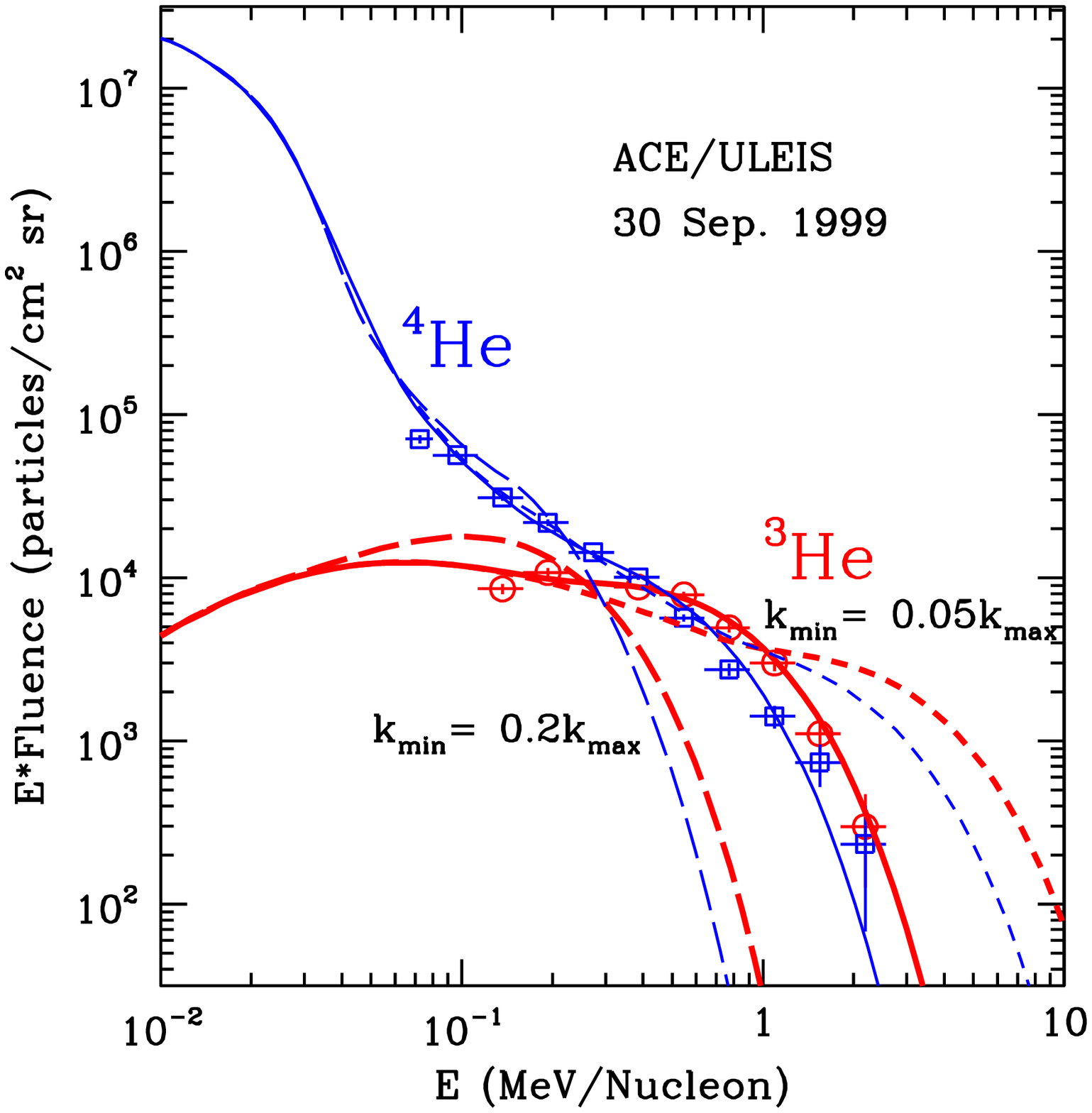}
\hspace{-0.6cm}
\includegraphics[height=5.0cm]{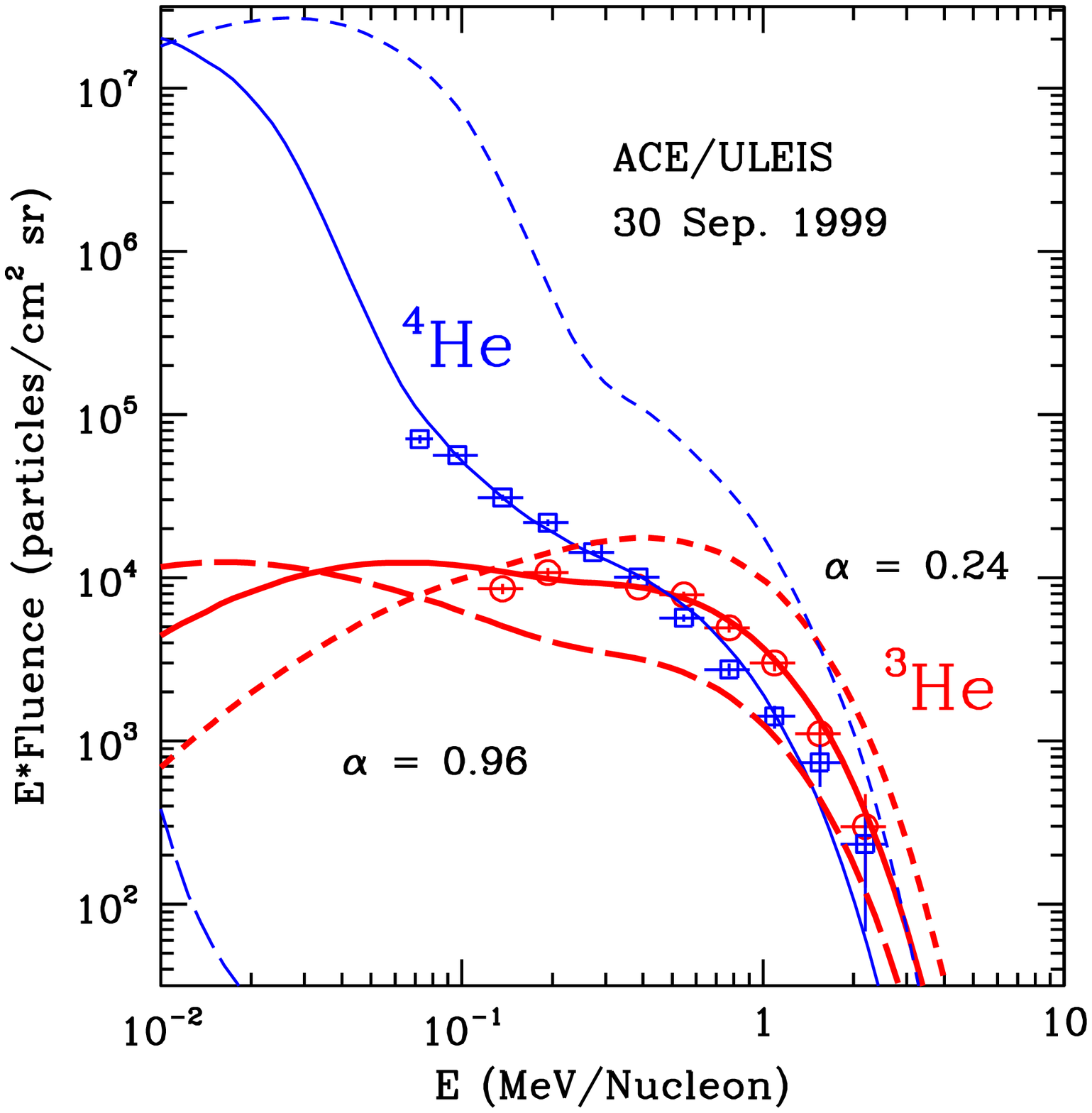}
\hspace{-0.6cm}
\includegraphics[height=5.0cm]{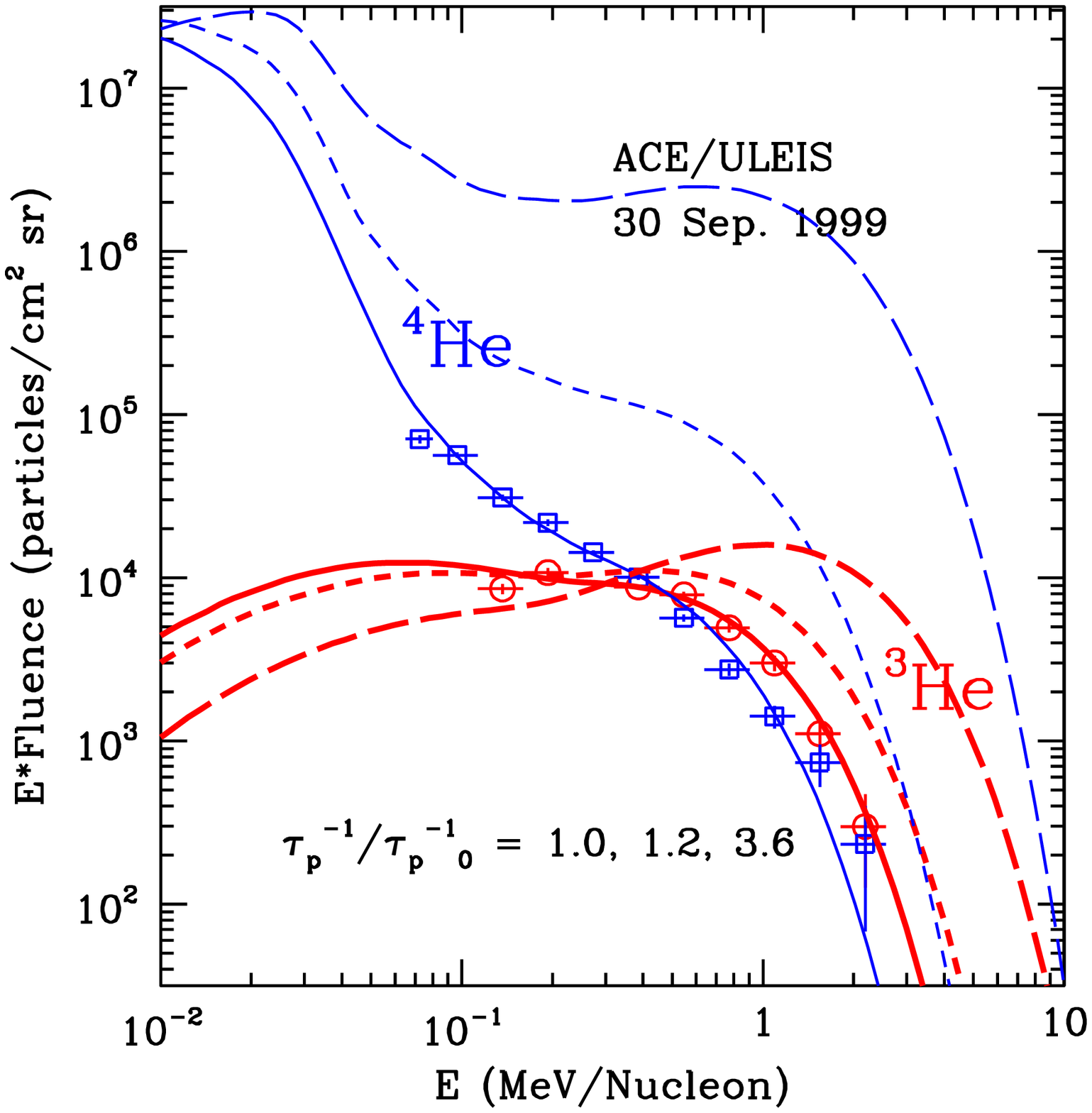}
\end{center}
\caption{\scriptsize
Dependence of the accelerated spectra of \he4 and \3he on $k_\min$ (left),
$\alpha$ (middle) and $\tau_p^{-1}$ (right; $\tau_p^{-1}$ in units of $\tau_p^{-1,0}=0.0055$
s$^{-1}$ ).  
The lines are labeled with the corresponding
numbers of each parameter.
In each case the  solid lines are for the  fiducial model  with $\alpha=0.5,
k_\max=2\alpha\delta^{-1/2}=10k_\min, q=2$ and $q_h=4$ that is chosen to fit the
data point shown for the 30 Sep. 1999 event observed by {\it ACE}. Note that for
a better
indication what energy particles dominate the spectra in a log-log plot we plot
particle energy time fluence.
}
\label{spectra}
\end{figure}

From the spectra we can calculate the ratio of \3he to \he4 fluences for
different models which could be then compared with the observed ratios shown in
Figure \ref{obs}. Inspection of observed spectra indicate that a
representative ion energy would be 1 MeV/nucleon. In Figure \ref{ratios} we
show the variation of this ratio with temperature (left panel) and  
$\tau_p^{-1}$ (middle and right panels) for several values of other important
parameters. As evident this ratio is most sensitive to the value of
$\tau_p^{-1}$ which represents the general rate of acceleration or the level of
turbulence. The ratio can change from the highest observed value  ($\sim 30$)
to  near photospheric value  ($\sim 2\times 10^{-4}$) for only a factor of 30
change in $\tau_p^{-1}$. It is natural to expect higher level of turbulence 
generation (i.e.  a larger value of $\tau_p^{-1 }$) in stronger  events.
Therefore, this predicted correlation is in agreement with the general trend of
observation shown in Figure \ref{obs} (left panel), if the strength of an event
is measured by the observed fluence of \he4 ions and most other ions like
carbon, nitrogen and oxygen.%
\footnote{It should be noted that while the observations are for fluences
integrated from 0.2 to 2.0 MeV/nucleon our theoretical ratios are calculated at
1 MeV/ nucleon  which is near the geometric or algebraic mean of the range.}  
This seems reasonable and calls for  more quantitative comparison with
observations and model prediction. In the next section we present one such
comparison.

\begin{figure}[htb]
\begin{center}
\includegraphics[height=5.0cm]{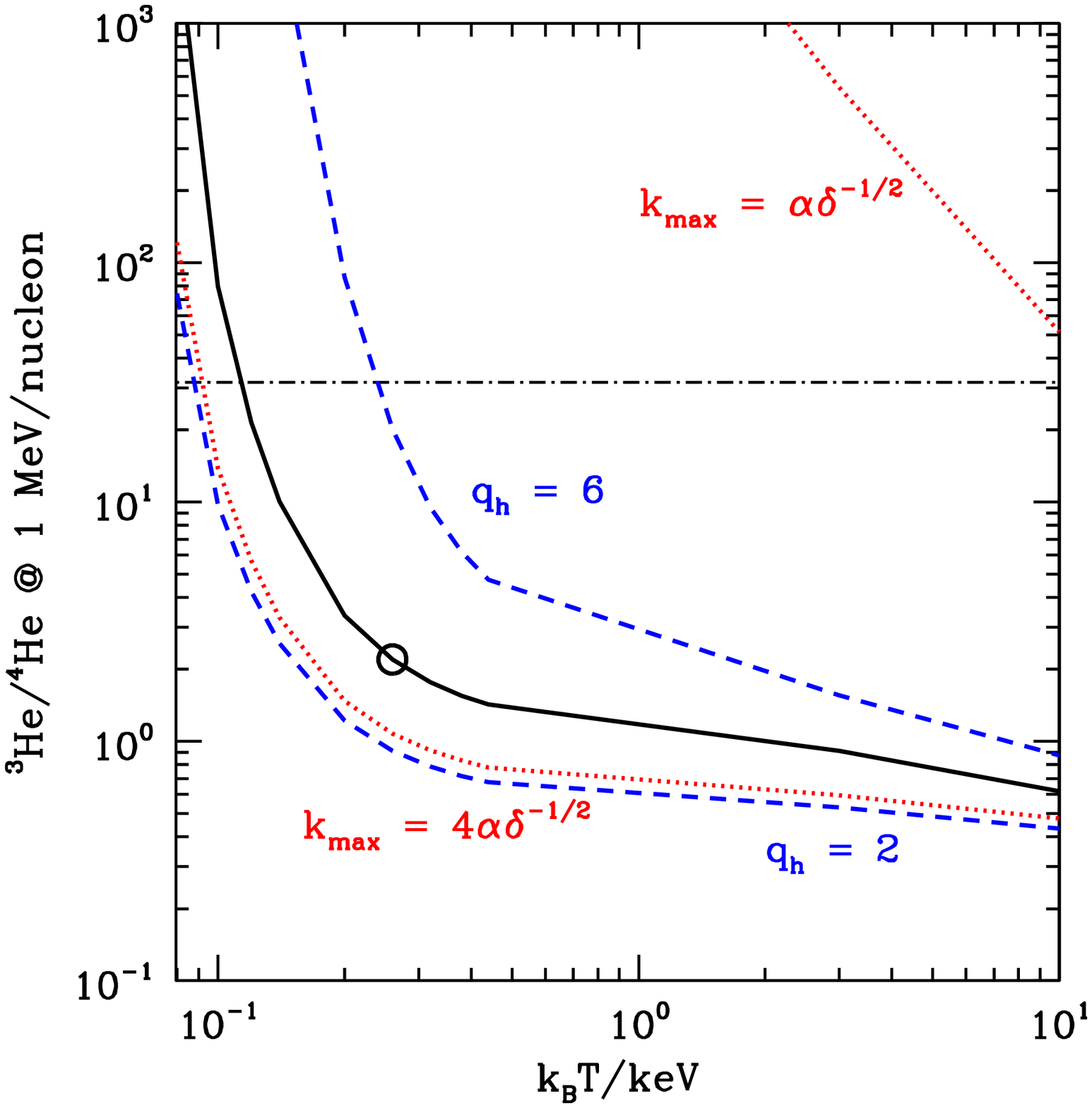}
\hspace{-0.6cm}
\includegraphics[height=5.0cm]{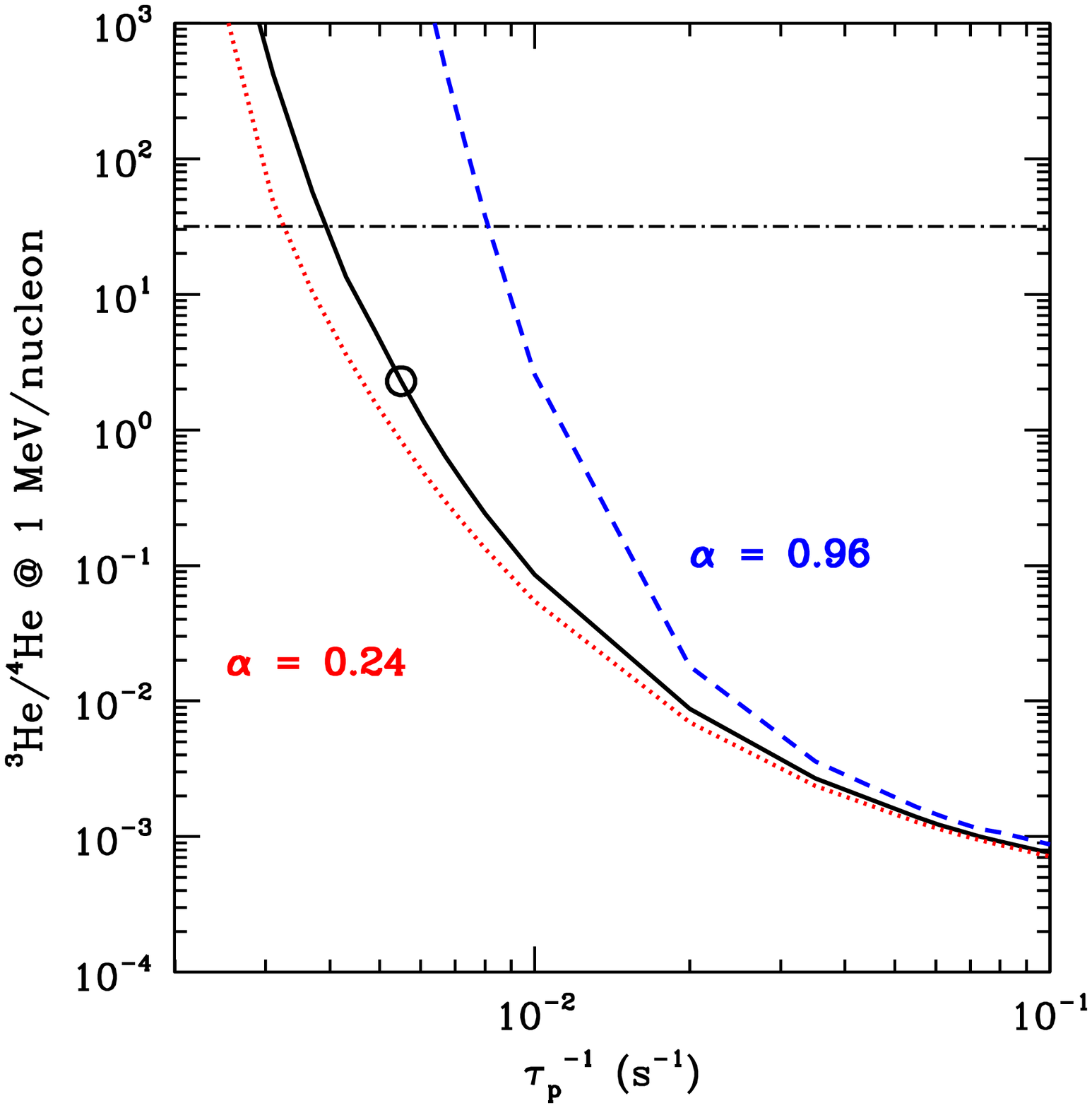}
\hspace{-0.6cm}
\includegraphics[height=5.0cm]{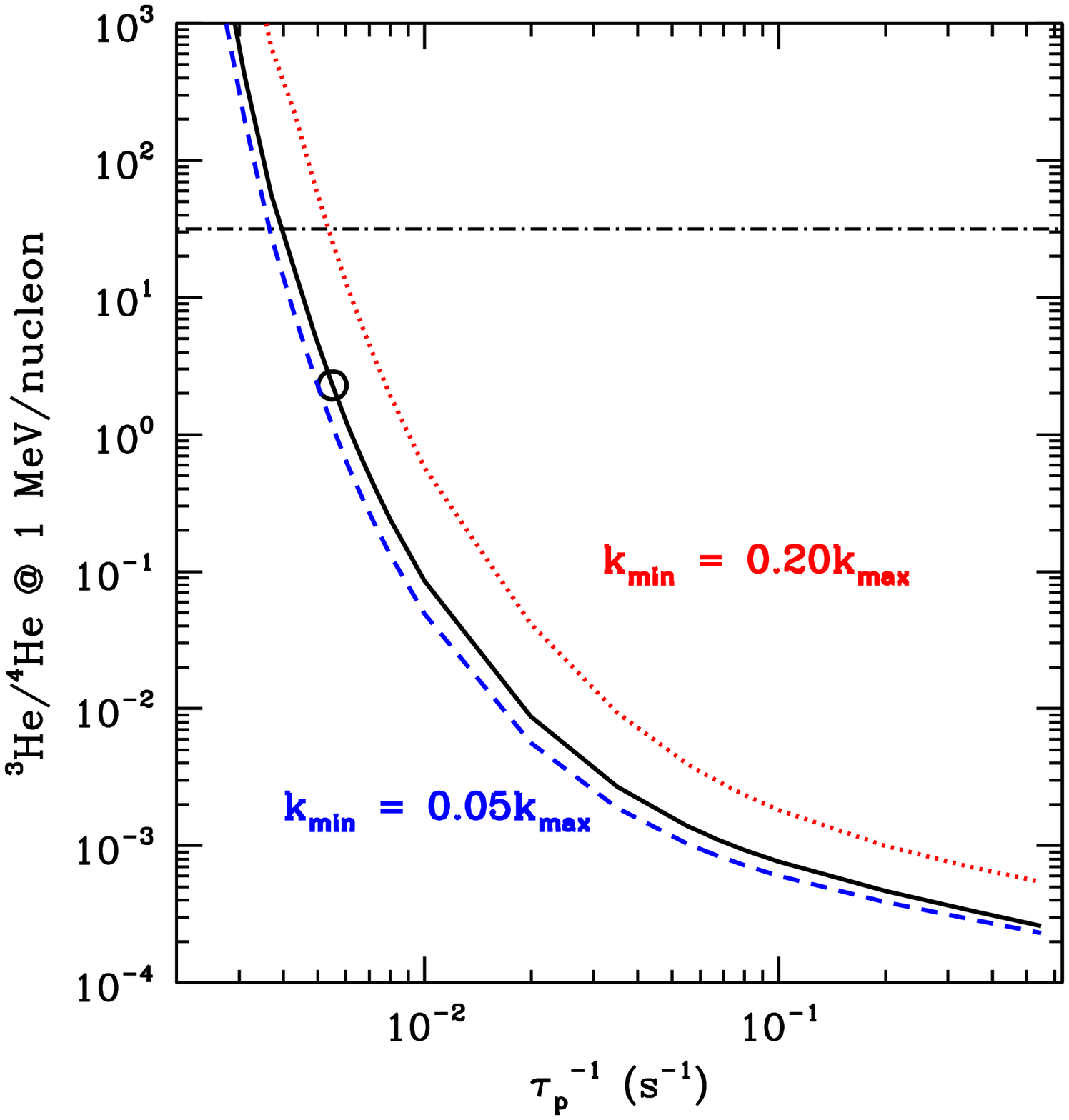}
\end{center}
\caption{\scriptsize
Variation  of the accelerated \3he to  \he4 fluence ratio (at $E=1$ MeV/nucleon)
with background plasma temperature $T$ (left) and
$\tau_p^{-1}$ (middle and right) for several values of other specified 
model parameters.  
The lines are labeled with the corresponding numbers of each parameter.
In each case the open circle stands for the model that fits spectra of  the 30
Sep. 1999 event, and the solid lines are for the  fiducial
model  with $\alpha=0.5, k_\max=2\alpha\delta^{-1/2}=10k_\min, q=2$ and $q_h=4$.
Note the weak dependence on the temperature for $T>2\times
10^6$ K and a strong dependence on $\tau_p^{-1}$ for all model parameters with
saturates at chromospheric values of the ratio. The horizontal dot-dash line
shows the highest ratio observed so far (see Fig. \ref{obs}, left). 
}
\label{ratios}
\end{figure}

\section{Distributions of Fluences}

\begin{figure}[htb]
\begin{center}
\includegraphics[height=5.2cm]{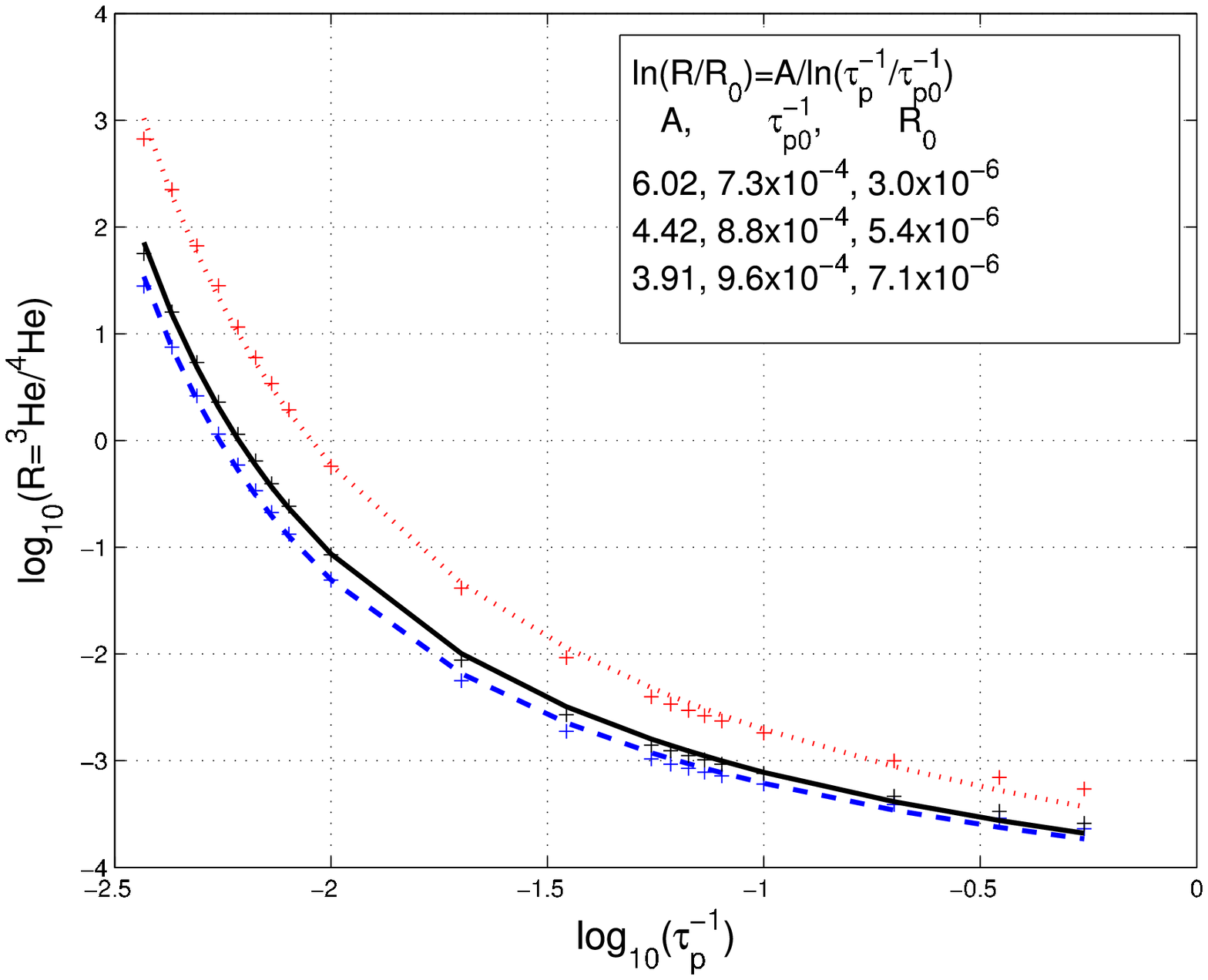}
\hspace{+0.6cm}
\includegraphics[height=5.2cm]{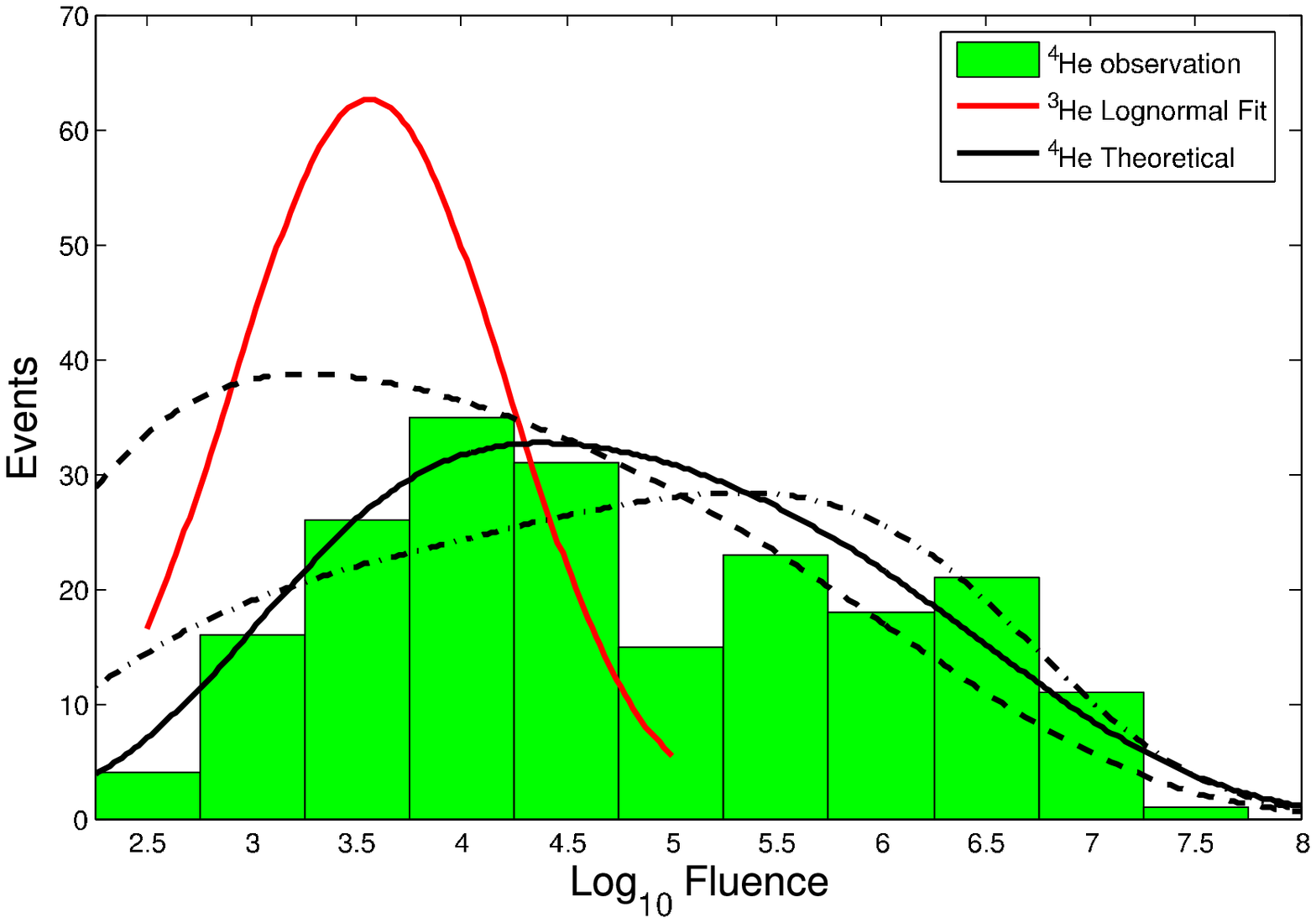}
\end{center}
\caption{\scriptsize {\it Left:} A simple analytic fit (curves) to the model
relations (points) between the fluence ratio and the acceleration rate or event
strength represented  by $\tau_p^{-1 }$    for the  three different values of
$k_\min$ of
the right panel of Figure \ref{ratios} with the indicated fitting parameters.
{\it Right:} The fitted log-normal distribution to the \3he fluences and 
predicted \he4 distributions of three models  compared with observations. The
solid line which gives the best fit is for  $n=2, k_\min=0.1k_\max$, the dashed
line is for $n=2, k_\min=0.2k_\max$  and the dash-dot line is for  $n=1.5,
k_\min=0.2k_\max$.
}
\label{results}
\end{figure}

We have seen that the general observed behavior of the the ratio of the fluences
defined as $R=F_3/F_4$  is similar to the model predictions. In this section we
try to put this result on a firmer quantitative footing by considering the
observed distributions of the fluences of both ions as shown in Figure \ref{obs}
(right panel).  Except for the minor truncation at high values of $F_4$,  the
fluence
of \he4, the observed distribution of $F_3$, the fluences of \3he, seem to
be almost bias free and not affected significantly by the observational
selection effects. For example, there are well defined and steep decline both at
the high and low fluences away from the peak \3he value of $F_0\sim 10^{3.7}
{{\rm 
particles} \over {{\rm  cm}^{2} {\rm sr} {\rm (MeV/nucleon)}}}$. This is not
what
one would expect if the
data suffered  truncation due to a low observation threshold. In such a case one
would observe a
distribution increasing up to the threshold followed by a rapid cutoff below it.
Our model results described above
also seem to predict the observed behavior. As stressed in previous
section, the \3he
spectra and fluxes appear to be fairly independent of model parameters because
essentially under all conditions most \3he ions are accelerated and form a
characteristic concave spectrum. Thus we believe that it is safe to assume that
the observed \3he distribution is a true representations of the intrinsic
distribution (as produced on the Sun).  This distribution can be fitted very
nicely with a log-normal expression.%
\footnote{The truncation shown by the shaded area in the middle panel of Figure
\ref{obs} introduces a slight bias against detection of low fluences. We
estimate
that, because there are fewer events at the high \he4 fluence end, this means a
10 to 20\% underestimation of the distribution of the three lowest values of the
\3he histogram (right panel, Fig. \ref{obs}). We will ignore this small
correction, whose main effect is to increase the value of $\sigma_3$ by a small
amount.} 
If we define the logs of the  fluences and their ratio as 
\beq
LF_3\equiv\ln (F_3/F_0),\,\,\,\,\  LF_4\equiv\ln (F_4/F_0) \,\,\,\,\ LR\equiv\ln
R,
\label{defs}
\eeq
then from fitting the observed distribution of \3he by a log-normal form we
get: 
\beq
\psi_3(LF_3)=\phi_0\exp\left({LF_3 \over \sigma_3}\right)^2\,\,\,\,\, {\rm
with}\,\,\,\,\, \sigma_3\sim 0.22,
\label{dist3}
\eeq
which is shown on the right panel of Figure \ref{results}. 

Using this distribution we now derive the distribution of \he4 fluences,
$\psi_4(LF_4)$. For this we use the model predicted relationship between the two
fluences as shown in Figure \ref{ratios} above. We will use the  two panels of
this
figure showing the dependence of the log of the fluence ratio $LR$ on 
$\tau_p^{-1 }$.
It turns out that most of these curves can be fitted by a simple function:
\beq
\ln (R/R_0)=LR-\ln R_0={A\over \ln (\tau_p^{-1 }/\tau_{p0}^{-1 })}.
\label{relation}
\eeq
The left panel of Figure \ref{results} shows fits to the curves in the right
panel of Figure \ref{ratios} with the indicated values of the the fitting
parameters $A, R_0$ and $\tau_{p0}^{-1 }$ (which is not the same as the 
$\tau_{p,0}^{-1 }=0.0055 {\rm s}^{-1}$ in Fig. \ref{spectra}). We
shall use this relation to transfer the \3he fluences and distributions to those
of  \he4.
 
For a given value of  $\tau_{p}^{-1 }$ the number of events with \he4 
log-fluences between  $LF_4$ and $LF_4+d(LF_4)$ (i.e. $\psi_4(LF_4)d(LF_4)$) is
equal to $\psi_3(LF_3)d(LF_3)$, the number of events with  \3he log-fluence
$LF_3 = LF_4 +LR(\tau_p^{-1})$ and  $LF_3+d(LF_3)$, where$d(LF_3)=d(LF_4)$and 
$LR(\tau_p^{-1})=\ln R_0 + A/\ln (\tau_p^{-1}/\tau_{p,0}^{-1 })$. Thus  we have
\beq
\psi_4(LF_4)=\psi_3(LF_4+LR[\tau_p^{-1}])=\phi_0\exp\left({LF_4+LR(\tau_p^{-1})
\over \sigma_3}\right)^2
\label{dist41}
\eeq
However, we expect not a single value for  $\tau_p^{-1 }$, which as stated above
is a proxy for the strength of
the event,  but a broad distribution of events with different strengths, say
$f(\tau_p^{-1 })$. Since, as argued above, the \3he fluence distribution
$\psi_3(LF_3)$ is independent of $\tau_p^{-1 }$,  then for a population of
events we have
\beq
\psi_4(LF_4)=\int_0^\infty \phi_0\exp\left({LF_4+LR(\tau_p^{-1}) \over
\sigma_3}\right)^2f(\tau_p^{-1 })d\tau_p^{-1 }.
\label{dist42}
\eeq
Every term in the above equations is determined by observations and our models
except the distribution $f(\tau_p^{-1 }$, which is a reflection of the level of
the distribution of the level of turbulence and, when multiplied by the volume
of the turbulent acceleration region (which does not affect the \3he/\he4
ratio), is related the overall strength of the event. 
Observations of solar flares show that most extensive  characteristics
which are a good measure of the flare strength or magnitude, such as X-ray,
optical or radio fluxes, appear to obey a steep power law distribution, usually
expressed as a cumulative distribution $\Phi(>F_i)\propto F_i^{-n}$  (or 
differential distribution $\phi(F_i)\propto F_i^{-n-1}$) with typically $n\sim
1.5$ (see, e.g.  Dennis 1985 and reference therein). Such a distribution seems
to roughly
agree with the prediction of the so-called avalanche  model proposed by Lu \&
Hamilton (1991).
Now assuming that  $\tau_p^{-1 }$ also obeys such a power law distribution
(\i.e. $f(\tau_p^{-1 })\propto (\tau_p^{-1 })^{-(n+1)}$) we can write the
distribution of \he4 as:
\beq
\psi_4(LF_4)=\int_0^\infty \phi_0\exp\left({LF_4+\ln R_0 +A/x \over
\sigma_3}\right)^2e^{-nx}dx, \,\,\,\,\, {\rm with }\,\,\,\,\,
x\equiv \ln (\tau_p^{-1 }/\tau_{p0}^{-1 }).
\label{dist43}
\eeq
Using the above relations we have calculated the \he4 fluence distribution. The
results for three models are compared with the observations on the right panel
of Figure \ref{results}. Given  the other model parameters ($k_\min, \alpha$
etc.) we have only one free parameter namely the index $n$ for this fit. The
solid line obtained for the top curve of the left panel ($k_\min=0.2k_\max,
\alpha=0.5$), and for $n=2$ provides a good fit to the observed distribution of
\he4 fluences.  In order to demonstrate the sensitivity of the results to the
parameters we also show two other model predictions based on slightly different
parameter values. These results provide additional quantitative evidence (beside
those given in LPM04 and LPM06) on the validity of the SA of SEPs by turbulence,
and  indicate that  with this kind of analysis one can begin to constrain model
parameters.

\section{SUMMARY AND CONCLUSIONS}

In this paper we have carried out further comparison between the prediction of
models based on stochastic acceleration of SEP ions by turbulence. In our
earlier works (LPM04, LPM06) we demonstrated that the extreme enrichments of
\3he and spectra of \he4 and \3he  observed in several events can be
naturally described in such a model. Using the results based on this model, here
we consider the relative distributions of \he4 and \3he fluences derived from a
large sample of event by Ho05. We show that with some simple and reasonable
assumptions we can explain the general features of these observations as well.

These are clearly preliminary results and are intended to demonstrate
that in addition to modeling only few bright events it is also important to look
at population as  a whole and ascertain that a model which can explain the
detail
characteristics of individual events can also agree with the distributions of
observables for a large sample of events. Here we have shown how the dispersion
in one parameter, namely the acceleration rate or the strength of the flare, can
account for the observed distributions of fluences. The key assumption here is
that the amount  of produced turbulence  (represented by $\tau_p^{-1}$) has a
wide dispersion and obeys a power law distribution similar to that observed for 
other extensive parameters that give a measure of the strength of a flare. 
The dispersion in
other model parameters can also influence the final outcome. However, the
dispersion of most of
the other important parameters, like  intensive parameters temperature, density
and 
magnetic field, are  expected to be much
smaller than that of an
extensive parameter like the overall strength of the event, the amount of
turbulence produced, the flare volume etc. In addition, as shown in the previous
section,
the intensive parameters play a lesser role than the extensive parameter
$\tau_p^{-1 }$ in determining the relative characteristics of \3he and \he4.
Given the dispersion of any other parameter one may carry out similar
integration
over its range. However, for the reasons given above we expect smaller changes
in the shapes of predicted distribution due to dispersion of most of the
intensive parameters. Given a more extensive set of data such improvements may
be needed can be  carried out.

The existing data may be used to test some of our assumptions, in
particular the assumption of constancy of the \3he distribution. We intend to
address these in future works. We can also make the above results more robust by
using model
fluences integrated over the same spectral range as the observations instead of
fluences at 1 MeV/nucleon. One can also expand this approach and address the
distributions of other characteristics besides the fluence, suc as the
spectral indexes or break energies (if any). The available data
contain this
information but require more analysis. 

{}

\end{document}